# Broadband omnidirectional invisibility for sound in three dimensions

Weiwei Kan[1], Bin Liang[1,2,*], Ruiqi Li[1], Xue Jiang[1], Xin-ye Zou[1,3], Lei-lei Yin[2], and Jianchun Cheng[1,3,*]

[1]Key Laboratory of Modern Acoustics, MOE, Institute of Acoustics, Department of Physics, Nanjing University, Nanjing 210093, P. R. China

[2]Imaging Technology Group, Beckman Institute, University of Illinois at Urbana-Champaign, Urbana, Illinois 61801, USA

[3]State Key Laboratory of Acoustics, Chinese Academy of Sciences, Beijing 100190, P. R. China

[*] Correspondence and requests for materials should be addressed to J. C. C. (email: jccheng@nju.edu.cn) or B. L. (email : liangbin@nju.edu.cn)

Acoustic cloaks that make object undetectable to sound waves have potential applications in a variety of scenarios and have received increasing interests recently. However, the experimental realization of a three-dimensional (3D) acoustic cloak that works within broad ranges of operating frequency and incident angle still remains a challenge despite the paramount importance for the practical application of cloaking devices. Here we report the design and experimental demonstration of the first 3D broadband cloak capable of cancelling the scattering field near curved surfaces. Unlike the ground cloaks that only work in the presence of a flat boundary, the proposed scheme can render the invisibility effect for an arbitrarily curved boundary. The designed cloak simply comprises homogeneous positive-index anisotropic materials, with parameters completely independent of either the cloaked object or the boundary. With the flexibility of applying to arbitrary boundaries and the potential of being extended to yield 3D acoustic illusion effects, our method may take major a step toward the application of acoustic cloaks in reality and open the avenue to build other acoustic devices with versatile functionalities.

Invisibility cloak, with the capability of making scattering objects undetectable from the EM signals [1-7] or acoustical signals[8-11], has become one of the most intriguing and investigated topics in the last decade. Among various methods developed for designing invisibility cloaks, the technique of coordinate transformation is a commonly-used as well as powerful one.[2,11-14] By coordinately transforming the physical fields into an illusion one where the space is filled with only background medium, the required material parameters for such device can be obtained. The parameters are often inhomogeneous, anisotropic, or of extreme values that do not exist in nature, until recently the advances in metamaterials make such parameters realizable.[15-21] Although numerous schemes have been proposed to design acoustic cloaks with practically feasible parameters[22-25], only a few experiments have demonstrated the cloaking effect for acoustic wave so far. Zhang et al. and Popa et al. claim the demonstration of free-space cloaking in water and ground cloaking in air respectively. And recently a three-dimensional (3D) axisymmetric cloak for a sphere in free-space based on scattering cancellation[10] is demonstrated by Sanchis et al.

In general, the variety kinds of cloaking can be divided into two categories:

free-space cloaking[8,10,11,14,26] and ground cloaking[5,7,9,27,28]. For free-space cloaks, the assumption of an infinite free space brings some difficulties or restrictions in its realization, and extreme parameters are often required which must be either negative[4] or very inhomogeneous[2,14]. As a result, free-space cloaks are usually limited to either a narrow bandwidth of working frequency or a certain range of view angle.[10,11,26] As another important class of cloaking shells, the ground cloaks are proposed for concealing objects positioned on a reflecting surface within a broad bandwidth. For EM waves, the experiment has been realized various times in this category and reached a certain level of maturity. Compared with the design of a ground cloak that only works near a totally flat boundary which is an ideal assumption, on the other hand, it should be more intriguing as well as challenging to hide an object near curved boundaries with arbitrary 3D geometrics which are apparently more common and representative in practice, e.g., the corner of a room or the slope of the seafloors, etc.[29] So far however, most of the experimental demonstrations for such cloaking [9,29] are preformed exclusively in two-dimensional (2D) geometries. This means these cloaks will be visible from the third dimension, heavily restricted from the application in reality. It still remains a challenge to experimentally realize a 3D broadband acoustic cloak that can cancel the scattering of incident wave from various directions, which should definitely take a major step toward the practical application of acoustic cloaking devices.

Here, we have proposed a design of a 3D cloak for objects near arbitrary curved surfaces on the basis of transformation acoustics. The designed cloak is composed by only anisotropic subwavelength structures and the required parameters are homogeneous-i.e., the parameters are invariant with position. Due to the absence of resonance element, the device can work in a broad bandwidth and is effective for any incident angle. The good performance of the fabricated design is characterized via measurements on the acoustic scattering from a cloak which is chosen, as a particular example, to make a round table in a bowl-like space undetectable from acoustic signals. This 3D acoustic cloak with simple design and easy fabrication constitutes a further significant step toward the real world application of cloaking and offer new possibilities of versatile manipulation on acoustic waves.

**Results**

For simplicity while without losing generality, we will demonstrate the performance of the proposed scheme by choosing a particular example as shown in Fig. 1. A small round table (diameter 12 cm and 7 cm height), which act as the object to be cloaked, is placed near a curved sound-hard surface that forms a space with bowl-like shape (15 cm depth). Note that both the object and the boundary have axisymmetric configurations, which is only for facilitating the fabrication of experimental sample as well as improving the precision of measurement. It should be stressed that the validation of our scheme is actually independent of the shape of either the object or the boundary.

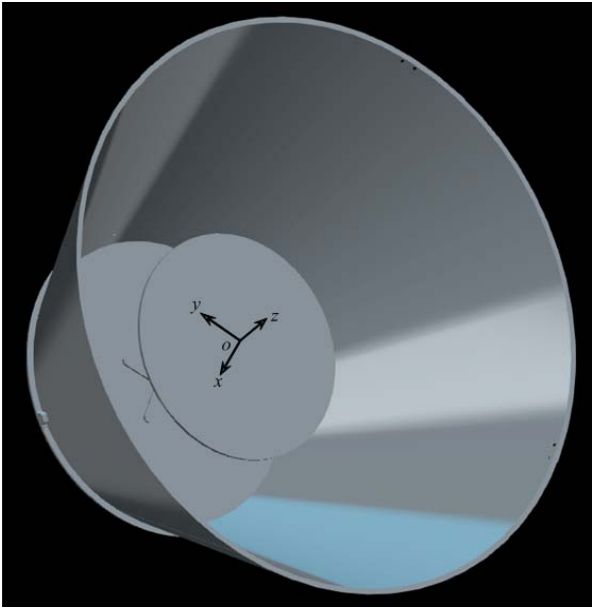

**Figure 1 | Schematic representation of the 3D object to be cloaked, chosen as a small round table placed in the bowl-like space.**

Our aim is to make the round table invisible from detection of acoustic signals and restore the original field in the bowl-like space by using a well-devised cloak covered onto the table. The fundamental design idea is acoustically squeezing the space near the surface along one direction (the direction along the $z$ axis in Fig. 1, e.g., the direction along the height of the table) into the cloak area. Then a blind space can be created where the objects can be hidden from the detecting signals. The ideal values for the required parameters of the anisotropic cloak medium are calculated as $0.47\rho_0$, $\frac{1}{0.47}\rho_0$, $0.47\kappa_0$ [see Methods], where $\rho_0$ and $\kappa_0$ are the mass density and bulk modulus of the background medium (air in this work) respectively. The performance of such cloak is numerically demonstrated in the symmetric plane ($y = 0$ plane) of the structure, as shown in Fig. 2.

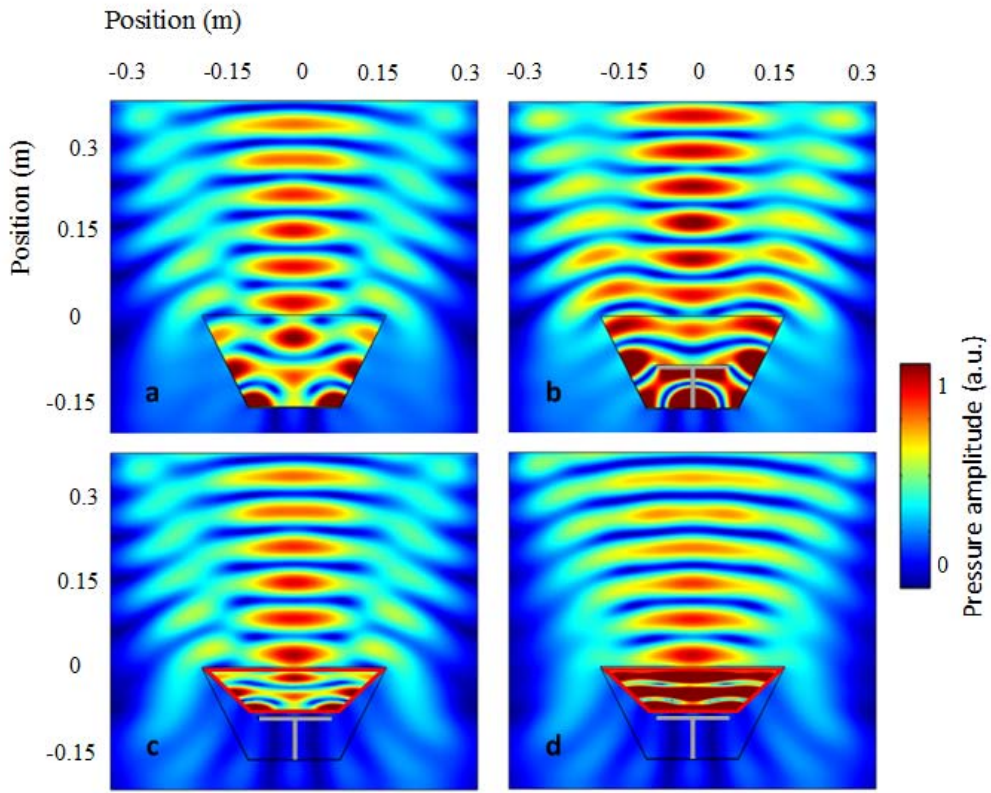

**Figure 2 | Numerical demonstrations in the symmetric plane of the structure, showing the effectiveness of the designed cloak.** Acoustic pressure amplitude fields (**a**) near the bowl-like sound hard boundary, (**b**) disturbed by the small round table, (**c**) for the table covered by the ideal cloak, (**d**) for the table covered by the reduced cloak. The silver object is the table and the proposed cloak is placed within the regions surrounded by red lines.

The 3.43 kHz plane wave is radiated from the upmost boundary and scattered by the structures. Other boundaries are assumed as matched boundaries or reflection-free boundaries, except the bowl-like sound hard surface. In Fig. 2a, there is nothing in the space surrounded by the bowl-like surface, so the plotted field is the pressure amplitude pattern for the total field (superposition of the incident wave field and the scattered wave field) near the bowl-like sound hard surface. In Fig. 2b, the round table is placed in the space, which affects the original field pattern obviously. Position shifts of certain minimum and maximum points in the total field can be clearly observed. After covering the round table with the devised cloak, the field pattern can be restored perfectly into the original one (see Fig. 2c), whereas the acoustic wave is only reflected by the hard boundary of the bowl-like surface. Because one component of the required effective density and the required bulk modulus is smaller than corresponding parameters of air, which suggests one basic material of the cloaking medium should be less dense and rigid than air. Such material is very difficult to find, then the eventual realizable parameters are enlarged with the same ratio and become $1.2\boldsymbol{\rho}_0$, $5.4\boldsymbol{\rho}_0$, $1.2\kappa_0$, for easy fabrication of the conceived device. Under this condition, the impedance of the effective media is modified, while the phase information is conserved. Therefore, there is some reflection at the boundary of the cloak and the background media, which do not exist for the ideal condition. But in the same time, the change of field pattern caused by the round table can still be reduced obviously, in other words, the reduced device can perform as an invisibility cloak as well. Such phenomenon can be observed in Fig. 2d clearly. Although the total acoustic field pattern is not strictly the same with that in Fig. 2a and c, it agrees better with Fig. 2a when compared with Fig. 2b. Especially, the positions of the local minimums and maximums of the total field pattern are restored. Even more, this design is valid for all incident angles, and the cloak effect can be even better for a certain incident angle, as the most inconformity of the acoustic impedance between cloak medium and background medium happens at the normal incidence and the reflection at the boundaries is the largest under this condition. Therefore, the cloak effect is mostly hampered under the normal incidence, as shown in Fig. 2d. When the incident angle becomes larger (defined as the angle between the wave vector and the normal of the boundary), the component of the incident wave traveling along the layers is larger, in which condition the impedance of the two medium matches better.

For realizing such parameters, we employed the subwavelength anisotropic unit cell shown in Fig. 3a, which can be easily made by drilling holes in a metal sheet. The geometric parameters determine the anisotropy and the effective acoustic parameters. By studying the normal incidence plane wave reflection and transmission of this unit cell, the geometry parameters are properly designed for the required effective mass density and bulk modulus. The thickness is 1 mm and the diameter of the hole is 1.6 mm. The cell size is chosen as 5 mm, which is 20 times smaller than the wavelength for frequency of 3430 Hz and sufficiently small to ensure the validation of the effective medium approximation. The retrieved effective parameters of this structure are calculated with the method proposed in Ref. [21] and shown in Figs. 3b and c as a function of frequency. The red line is the normalized effective mass density and the blue line is the normalized bulk modulus. It can be observed for Fig. 3b that in the low frequency region, from 0 Hz to about 4 kHz, the effective parameters nearly remains constant, which implies the resulting device is able to work in this broad band region. Even more, the band width can be broaden further by simply scaling down the unit cell sizes. Compared with local resonant structures, whose working frequency inevitably varies with the dimension of the unit cells, the lower bound of the working frequency of this anisotropic structure do not change, while the upper bound can be promoted if the cell size is scaled down and the viscosity effect can be neglected. While the frequency is increased to above 4 kHz, the structure will behave more like a sonic crystal, dispersion phenomenon emerges and the effective parameters no longer remain constant. For the 3D structure, the diagonalized mass density tensor should have three components. But for the symmetry of this structure, the components of density tensor in the two orthogonal directions along the plate are the same. So that we have only two independent components as we plotted along these two different directions in Figs. 3b and 3c respectively. Figure 3b is for the case that the wave normally incidents onto the plate, whereas the effective density is much larger than the case in Fig. 3a in which the incident wave goes along the plate. Such a structure will increase the momentum in the direction vertical to the plates, while affect little in the direction along the plates. We arrange the structure periodically in the cloak area; the whole structure actually forms thin layers of perforated metal plates.

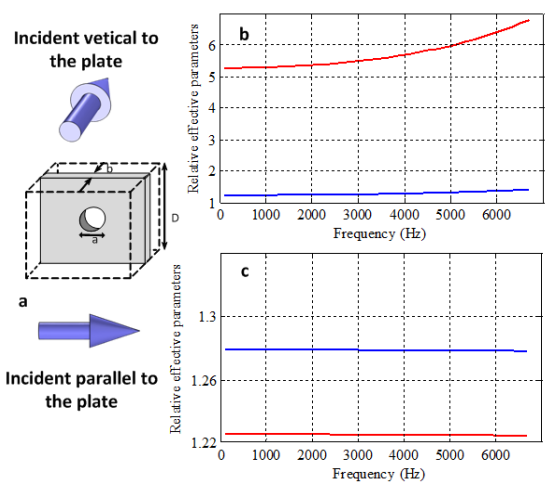

**Figure 3 | The effective parameters of the anisotropic structure.** (**a**) The illustration of the unit cell. (**b-c**) The retrieved effective parameters of this structure as

a function of frequency when the wave normal incidents to the plate (**b**) and goes along the plate (**c**). The red line is the normalized effective mass density and the blue line is the normalized bulk modulus.

The perforated metal plates are achieved by punching holes periodically in the thin metal plates, and then the whole structure was fabricated by assembling the plates with a properly designed shell molded by a commercial 3D printer. Figure 4a is the schematic of experimental setup, along with the photo of the setup (Fig. 4b) and prototype (Fig. 4c and d) . In the measurement, we simply put the loudspeaker 1 m away from the structures to mimic the plane wave field. Considering the fact that we test the performance of the device under the 3.43 kHz signals and the corresponding wavelength is about only 0.1 m, this can be regarded a considerably accurate approximation. Measurements were done in an anechoic chamber. The acoustic pressure amplitude field corresponds to the bowl-like surface, the bowl-like surface with the round table, the bowl-like surface with the round table and the fabricated invisibility cloak, is measured in the mapped region. For the axisymmetric configurations of both the object and the boundary, the field in the plane $y=0$ gives the complete information in the 3D space, so the measurements were only done in this plane for simplicity and clear display, and due to the limitation of the moving range of our linear stages and the symmetry of the measured systems, only the right side in Fig. 4a is studied in the experiment. Although the proposed scheme applies to arbitrary curved surfaces with any scattering objects, and under the detection of any acoustic signals, we only take the bowl-like surface, round table as an experimental verification for easy fabrication.

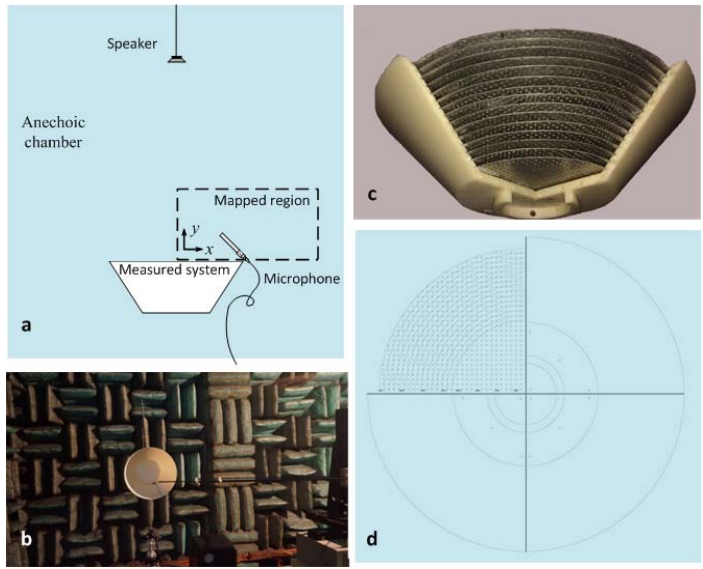

**Figure 4 | Experimental setup:** (**a**) the schematic of the experimental setups used to map the acoustic field near the cloak. The mapped region is indicated together with the sound source. (**b-c**) The photo of the whole measuring system (**b**) and the prototype of cloak (**c**). One side of the shell is taken off to show the interior structure. (**d**) Top view of the prototype.

Because the dimension of the scattering object (around 10 cm) is only of the same order of the detecting wave length, the scattering wave is not strong enough for obvious distinction among different wave patterns. As shown in Figs. 2 a and b, there exist some seeming similarities between the scattering fields generated by cloaked and uncloaked objects. For unambiguously evaluating the performance of the designed cloak, we used the parameter of field disparity (FD) defined as $\mathrm{FD} = |p(x,z) - p_0(x,z)|$, where $p(x,z)$ is the sound pressure amplitude measured at $(x,0,z)$ in the mapping region when a particular target is placed near the curved

surface, and $p_0(x,z)$ is the sound pressure amplitude measured at $(x,0,z)$ in the absence of the object. FD gives a quantitative estimation on the scattering ability of the target object. Vanishing of FD demonstrates cancellation of the scattering by the target object, i.e., the invisibility cloak functions well. Figure 5 illustrates the comparison of the measured FD for two cases where the target object is chosen as the round table [Fig. 5b] and round table covered with the invisibility cloak [Fig. 5a] respectively, and also the corresponding numerical simulations [Figs. 5c and d]. A good agreement is observed between the numerical prediction and the experimental result. It is apparent that the acoustic scattering field generated by the scattering object alone is significantly different from that of the invisibility illusion, as shown by the large fluctuations in Figs. 5b and d. Contrarily, both the numerical and experimental results of the FDs are negligible (see Figs. 5a and c) when covered with the cloak, indicating that the round table is made invisible from detecting signals by the cloak device, e.g., the existence of the cloak makes the space with the round table to be acoustically perceived as the illusion that nothing is there. As the invisibility cloak actually functions by creating an illusion filled with only the background media, the scheme can easily be extended to design an illusion cloak by choosing a different illusion space, instead of the illusion with only the background media. Under the detection of outside signals, the hidden object will be acoustically transformed in to the freely-designed illusion by the 3D illusion cloak. The required parameters will be similarly devised (see Methods) and the fabricating complexity only depends on the complexity of the acoustical parameter distribution of the chosen illusion space.

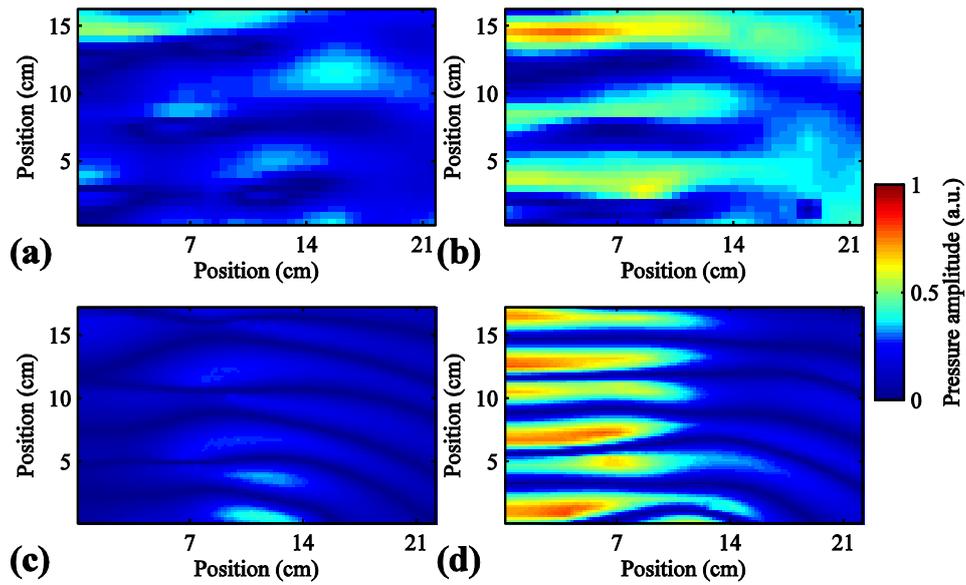

**Figure 5 | Experimental results.** Experimental results for the cloak with its corresponding numerical result. The field of FD in the $y=0$ plane when the target is chosen as (**a**) the table without the cloak, and (**b**) the table with the cloak placed in the bowl-like space. (**c-d**) Corresponding numerical results.

The sound source in the experiment is adopted as a 5 cm diameter loudspeaker, with a specific directional pattern. As the measured system and mapped area are in a small space angle, the wave field was verified as a good plane wave field in the mapped area before setting up the measured system. Besides, the sound attenuation was not considered in this work, although the attenuation, especially in the cloaking medium, may bring some influences to the result and will be addressed in the future research. These conditions may lead to the difference between patterns of FDs in simulation and experiment, but do not obviously impair the performance of the invisibility cloak.

This general scheme of acoustic cloaking have no specific restrictions, as the device can work in broad bandwidth and omni-directionally, the parameters are independent with the cloaked object and the geometry of the boundaries. Besides, it has no restriction to the size of the cloaked objects in theoretical. But some factors should be taken in to consideration for proper design in specific conditions. For example, it is still difficult to hide a large object with a small sized cloak. This requires a high relative refractive index in the compressing direction for the cloaking medium, and will cause strong mismatch between the cloaking medium and air as the realizable parameters for the cloaking medium are usually obtained by multiply the ideal value with a constant. Although the scaling of parameters can be avoided for cloaking in water, generating a high relative refractive index material in water is difficult in itself. When there is enough space for the cloaking device, increasing its size will help to improve the cloak effect. Although it is theoretically possible to make a huge cloak for hiding a big object from high frequency detecting signals, the fabrication will be a heavy work as the device would be composed by millions of unit cells, of which the dimension should be much smaller than the wavelength in the background medium.

**Discussion**

We reported the first experimental demonstration of 3D acoustic cloak for objects near arbitrary curved surfaces by exploiting the technique of transformation acoustics. and experimentally verified the scheme by making a small round table in a bowl-like space undetectable by acoustic waves. The cloak medium is only composed by homogeneous and anisotropic metamaterial, of which the parameters are easily realized by subwavelength structures. Moreover, the parameters of the cloak shell are independent of the properties of either the original object or cloaked region, which brings significant convenience for the design of such devices. The fabricated 3D cloak is able to function in a broad bandwidth and for all incident angles of detecting signals. All these feathers will significantly facilitate the practical application for such devices.

**Methods**

**Theory.** The device is designed according to transformation acoustics, the relationship of material parameters between virtual and physical systems can be written as $\boldsymbol{\rho}'_1 = \boldsymbol{H} \boldsymbol{\rho} \boldsymbol{H}^{-1} \boldsymbol{\tilde{H}}/\det(\ )$, and $\kappa'_1 = \det(\boldsymbol{H}) \times \kappa_1$, in which $\boldsymbol{H}$ is the Jacobian defined as $\boldsymbol{H} = \text{diag}\{\partial x'_1/\partial x_1,\ \partial y'_1/\partial y_1,\ \partial z'_1/\partial z_1\}$. As the physical space can be regarded as a compressed virtual space, the mapping between the physical region and the virtual one is specified by $x'_1 = x_1$, $y'_1 = y_1$ and $z'_1 = z_1/n_c$, where $n_c$ is the index of compression, and take a value of $1/0.47$ in this work. Then the parameters of the cloak are

$$\boldsymbol{\rho}'(x, y, z) = \text{diag}\{1/n_c,\ 1/n_c,\ n_c\} \rho_v(x, y, z),$$

$$\kappa'(x, y, z) = \kappa_v(x, y, z)/n_c.$$

**Numerical simulation.** The numerical simulation in Fig. 2 is performed with general PDE module of COMSOL Multiphysics® (the commercial software package based on finite-element method).

**Acoustic Measurement**. The experiment is carried out inside an anechoic chamber,

the setups and acoustic field mapping area is illustrated in Fig. 4. A 5cm-diameter sound speaker is used to excite the approximate plane wave field around the structures with a sinusoidal signal. Considering the difficulty in fabricating the cloak devices with subwavelength structure at high frequency region, the measurement is carried out under a relatively low driving frequency of 3.43 kHz to guarantee the precision of the structural parameters of sample. The sound pressure is measured by 0.25-inch-diameter Brüel & Kjær Type-4961 microphones. The microphone is fixed on to a linear stage with graphite shaft and can move in the mapping region with high accuracy. The recording and analysis equipment contain a Brüel & Kjær PULSE 3160-A-042 multichannel analyzer and a desktop computer with Brüel & Kjær PULSE software LabShop version 13.5.10.

**Acknowledgements**

This work was supported by the National Basic Research Program of China (973 Program) (Grant Nos. 2010CB327803 and 2012CB921504), National Natural Science Foundation of China (Grant Nos. 11174138, 11174139, 11222442, 81127901, and 11274168), NCET-12-0254, and A Project Funded by the Priority Academic Program Development of Jiangsu Higher Education Institutions. We acknowledge useful discussions with Jose Sanchez-Dehesa.

**Author contribution statement**

W. W. K., R. Q. L., X. J. and X. Y. Z. conducted the theoretical simulations and experiments. L. L. Y. fabricated the device. W. W. K. and B. L. interpreted the data. W. W. K., B. L. and R. Q. L. constructed the experimental set-up. J. C. C. and B. L. conceived and supervised the study. W. W. K., B. L., and J. C. C. wrote the paper.

**Additional information**

Competing financial interests: The authors declare no competing financial interests.

**Figure legends**

**Figure 1 | Schematic representation of the 3D object to be cloaked, chosen as a small round table placed in the bowl-like space.**

**Figure 2 | Numerical demonstrations in the symmetric plane of the structure, showing the effectiveness of the designed cloak.** Acoustic pressure amplitude fields (**a**) near the bowl-like sound hard boundary, (**b**) disturbed by the small round table, (**c**) for the table covered by the ideal cloak, (**d**) for the table covered by the reduced cloak. The silver object is the table and the proposed cloak is placed within the regions surrounded by red lines.

**Figure 3 | The effective parameters of the anisotropic structure.** (**a**) The illustration of the unit cell. (**b-c**) The retrieved effective parameters of this structure as a function of frequency when the wave normal incidents to the plate (**b**) and goes along the plate (**c**). The red line is the normalized effective mass density and the blue line is the normalized bulk modulus.

**Figure 4 | Experimental setup:** (**a**) the schematic of the experimental setups used to map the acoustic field near the cloak. The mapped region is indicated together with the sound source. (**b-c**) The photo of the whole measuring system (**b**) and the prototype of cloak (**c**). One side of the shell is taken off to show the interior structure. (**d**) Top view of the prototype.

**Figure 5 | Experimental results.** Experimental results for the cloak with its

corresponding numerical result. The field of FD in the $y=0$ plane when the target is chosen as (**a**) the table without the cloak, and (**b**) the table with the cloak placed in the bowl-like space. (**c-d**) Corresponding numerical results.